\documentclass[prl,twocolumn,amsmath,amssymb]{revtex4-1}
\usepackage{graphicx}
\usepackage{bm,times}
\usepackage[colorlinks,citecolor=blue,linkcolor=red]{hyperref}
\usepackage{color}
\usepackage{comment}
\usepackage{lipsum}
\usepackage{braket}
\usepackage{scalerel}

\begin{document}
\title{Cycle Flux Ranking of Network Analysis in Quantum Thermal Devices}

\author{Luqin Wang$^{1}$}
\author{Zi Wang$^{1}$}
\author{Chen Wang$^{2}$}\email{wangchenyifang@gmail.com}
\author{Jie Ren$^{1}$}\email{Xonics@tongji.edu.cn}
\address{$^{1}$Center for Phononics and Thermal Energy Science, China-EU Joint Lab on Nanophononics,
Shanghai Key Laboratory of Special Artificial Microstructure Materials and Technology,
School of Physics Sciences and Engineering, Tongji University, Shanghai 200092, China}
\address{$^{2}$Department of Physics, Zhejiang Normal University, Jinhua 321004, Zhejiang, P. R. China
}

\date{\today}
\begin{abstract}
Manipulating quantum thermal transport relies on uncovering the principle working cycles of quantum devices. Here, we apply the cycle flux ranking of network analysis to nonequilibrium thermal devices described by graphs of quantum state transitions.
To excavate the principal mechanism out of complex transport behaviors, we decompose the quantum-transition network into cycles,
calculate the cycle flux by algebraic graph theory,
and pick out the dominant cycles with top-ranked fluxes, i.e., the cycle trajectories with highest probabilities.
We demonstrate the cycle flux ranking in typical quantum device models, such as a thermal-drag spin-Seebeck pump, and a quantum thermal transistor as thermal switch or heat amplifier.
The dominant cycle trajectories indeed elucidate the principal working mechanisms of those quantum devices.
The cycle flux analysis provides an alternative perspective that naturally describes the working cycle corresponding to the main functionality of quantum thermal devices, which would further guide the device optimization with desired performance.
\end{abstract}

\maketitle
\emph{Introduction}.
Harnessing heat and information in low dimensional nanoscale systems is overwhelming in the active research of
functional thermal devices~\cite{bli2004prl,cwchang2006science, karl2014apl,pba2013apl,sb2011prl,bli2006apl,lwang2007prl,lwang2008prl,Guarcello2018prapp,Kubytskyi2014prl,Klaers2019prl}, 
in analog with modern electronics~\cite{nbl2012RMP}.
The dramatic advance of nanotechnology further accelerates the demand for managing heat in quantum regime~\cite{gchen2005book},
which becomes prominent due to the ubiquitous quantum effect in nano-devices~\cite{nanores2010}.
Consequently, various quantum thermal devices modeled with dissipative open quantum systems are attracting tremendous attention~\cite{nitzan2005prl,truokola2011prb,rj2013prbndtc,kjoulian2017zna,pereira2019pre,Karg?2019pre,karl2017pre,bqguo2018pre,cwang2018pra,bqguo2019pre,jydu2019pre,yczhang2017apl}, paving the avenue for the realistic design of on-chip thermal devices. 

Recently, network theory becomes an important method to describe the quantum transport ~\cite{gaugefeild,prldisorder,manzano2013quantum,manzano2013quantum,prl115,prl10qubit,Ehrhardteabc5266}, where the nonequilibrium effects of external dissipative reservoirs are incorporated into the edge weights of the directed network of state transitions~\cite{cjs2013jcp}. In general, the network analysis of nonequilibrium transport can be established by cycle decomposition in the graph theory~\cite{schnakenbergRMP,andrieux2007fluctuation,Vollmerpre}, which leads to a flurry of inspiring works. 
For example, Nitzan's group investigated quantum transport control from the cycle veiwpoint in both photovoltaics' state networks~\cite{nitzan2014jpcc, nitzan2016jcp} and electrothermal networks~\cite{gtcraven2017prl}. J. Wang's group studied the anti-symmetric driving force via curl flux in the closed-loop of the quantum donor-acceptor model~\cite{zzd2015njp}. While considering the dynamical driving,
Jarzynski's group~\cite{jarzynskiprl}, and Chernyak and Sinitsyn~\cite{chernyakprl}, together with subsequent research~\cite{jarzynskijcm,asbanprl,Ren_2011}, revealed and emphasized the peculiar role played by cycle graphs in general no-pumping theorems of transports. Furthermore, as a kind of quantum machines, the full functioning of a quantum thermal device fundamentally requires completing cycles in the state space, as same as thermodynamic cycles in quantum heat engines~\cite{quanhaitaopre12007, quantummachine,paoloprl2020,prb2021} and quantum refrigerators~\cite{venturelli2013prl,prl105}. 
Therefore, the cycle graph analysis should be a powerful approach to depict the detailed mechanism of dissipative quantum thermal devices.

Nonequilibrium quantum transports are microscopically represented by directed cycle trajectories, which are formed of self-avoid closed quantum transitions. 
Such novel cycle trajectories are the key ingredients affecting the performance of quantum thermal devices. However, the formidable complexity is dictated by the huge number of cycle trajectories, composed of massive quantum states and transitions among them. 
This inspires us to explore the complex transport behaviors from the top-ranked cycle trajectories with largest occurring probabilities, i.e., cycle fluxes. 

In this Letter, 
by mapping quantum thermal devices into networks of quantum state transitions, we decompose the transition network into cycle trajectories, calculate and rank the cycle flux with the help of algebraic graph theory. By picking out the top-ranked dominant cycles, we could surmount the challenge caused by huge dimensional parameter space, to genuinely grasp the physical essence of the multi-functional quantum devices.
Concretely, we utilize two typical quantum device models to demonstrate the power of cycle flux and graph theory analysis. The dominant cycle fluxes are unraveled to dissect the principal mechanism out of complex transport features, such as thermal-drag spin-Seebeck pump, quantum thermal switch, heat amplification, and negative differential thermal conductance (NDTC).

\begin{figure}[htb]
\includegraphics[scale=0.23]{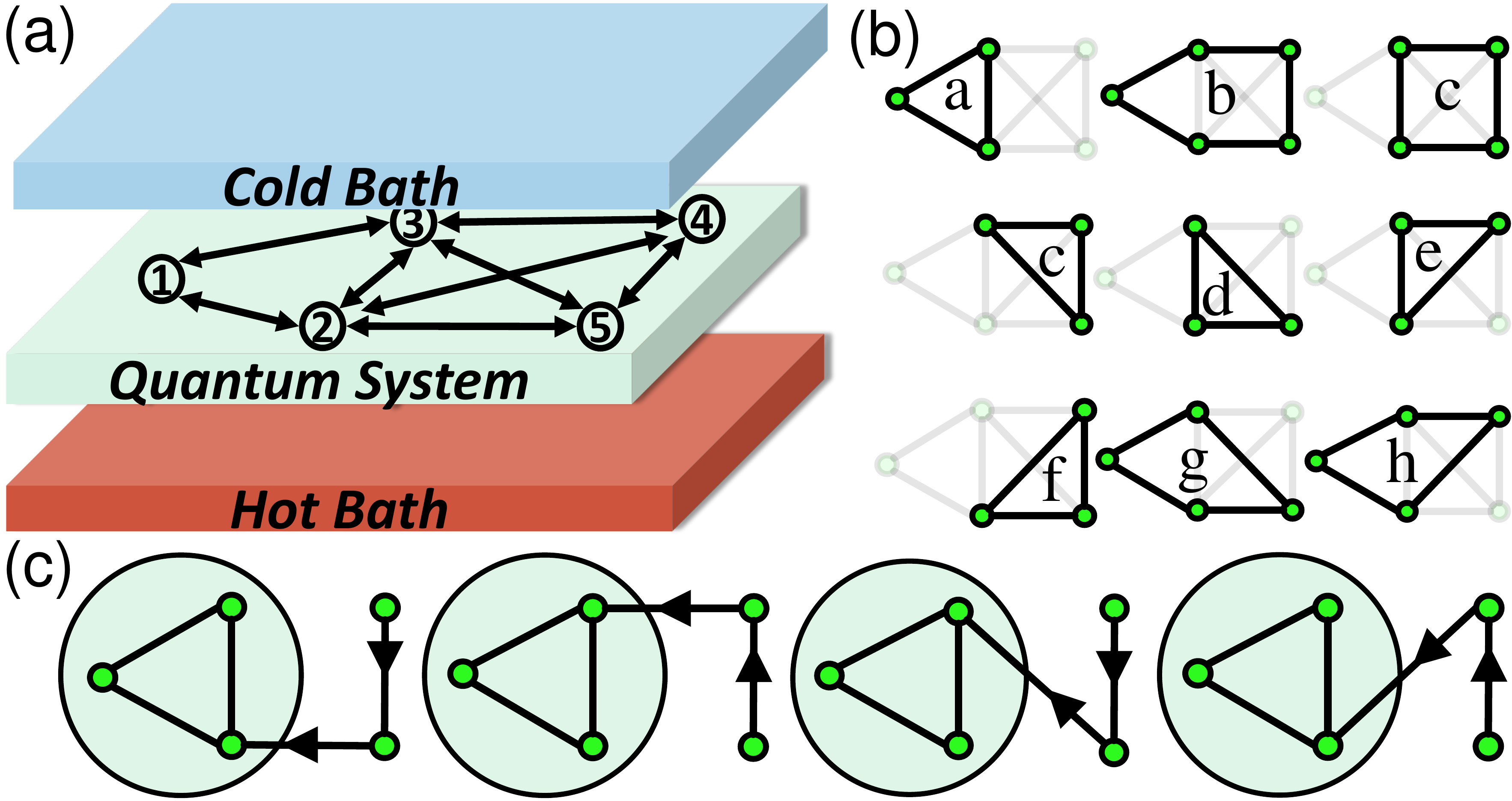}
\caption{\textbf{Mapping dissipative open quantum devices into networks.}
(a) Schematic of a transition network for quantum thermal transport under nonequilibrium conditions.
(b) All undirected cycle graphs contained in the transition network.
(c) All the spanning trees rooted on (every edge is oriented towards) cycle $C_{1\rightarrow2\rightarrow3\rightarrow1}$. 
}~\label{fig1}
\end{figure}

\emph{Connecting Quantum Transport with Cycle Graph Theory}.
The dissipative dynamics of open quantum systems are conventionally described by the quantum master equation in Lindblad form:~\cite{uweiss2008book,hp2002book}
\begin{equation}
\frac{d\rho_s(t)}{dt}=i[\rho_s,H_s]+\sum_v\mathcal{D}_v[\rho_s],
\label{lindbladeqn}
\end{equation}
where $\rho_s$ is the reduced density matrix, and $\mathcal{D}_v[\rho_s]$ is the Lindblad superoperator describing the dissipation induced by the $v$-th thermal reservoir (See Supplement section \uppercase\expandafter{\romannumeral1}~\cite{Suppl}). 
When fully thermalized by external reservoirs, the off-diagonal coherence $\rho_{ij}$ vanishes to zero when approaching steady states. Also, it was pointed out by Zurek~\cite{zurek1981pointer} that open quantum systems interacting with reservoir Hamiltonian can be cast into a diagonal form, with a wise choice of ``pointer basis". Moreover, Cao's group showed that by applying higher-order kinetic expansion, the quantum dissipative dynamics with coherence can be mapped into kinetic networks~\cite{cjs2013jcp}.
Therefore, without loss of generality, the quantum Lindblad equation can be extracted into a Pauli master equation:
$\frac{d}{dt}\rho_{ii}=\sum_{j}J_{j\rightarrow i}=-\sum_{j}L_{ij}\rho_{jj}$,  
where $\rho_{ii}$ denotes the probability to observe the system in state $\ket{i}$, and $J_{j\rightarrow i}=\sum_v k_{j,i}^v\rho_{jj}-k_{i,j}^v\rho_{ii}$ is the net edge flux from $\ket{j}$ to $\ket{i}$ with forward(backward) transition rate $k_{j,i}^v (k_{i,j}^v)$ induced by the $v$-th reservoir (See Supplement section \uppercase\expandafter{\romannumeral1}~\cite{Suppl}). $L_{ij}=-\sum_{v}k_{j,i}^{v}+\delta_{ij}\sum_{v,m}k_{j,m}^{v}$ is the Laplacian transition matrix describing the open dissipative system under nonequilibrium conditions. In the steady states, the edge flux has zero divergence due to the probability conservation so that $\sum_{i} L_{ij}=0$. As such, the nonequilibrium open quantum dynamics is represented by a graph with Laplacian matrix $L_{ij}$, as depicted in Fig.~\ref{fig1}(a), where the vertices denote the quantum states and the weighted directed edges denote the transitions from $\ket j$ to $\ket i$ with positive rates.

Alternatively, the graph can be represented on the basis of cycles and the edge flux can be decomposed into cycle fluxes, $ J_{i\rightarrow j}=\sum_{C}J_{C_+}-J_{C_-}$, where the summation is over all cycle trajectories $C$ that contain the edge $(i\rightarrow j)$, with subscript $+(-)$ for the forward (backward) cycle along (opposite to) $\ket{i}\rightarrow \ket{j}$ direction.  For example, Fig.~\ref{fig1}(b) lists all the cycles of the graph, and the net edge flux 
$J_{3\rightarrow 2}=(J_{a_+}+J_{c_+}+J_{d_+}+J_{e_+})-(J_{a_-}+J_{c_-}+J_{d_-}+J_{e_-})$. 
The cycle fluxes, pioneered by Hill~\cite{hill1966,hill1975} and developed by Kohler~\cite{kohler1980} and Schnakenberg~\cite{schnakenbergRMP}, can be intuitively understood as the cycle frequency, quantifying how many rounds the state can transit through one complete cycle per unit time, and have been widely carried out in describing biochemical systems~\cite{qianhong,wangjin,jren2017fop,udo2015prl,qian2016}. 

The cycle flux of the cycle trajectory $C_\pm$ is formulated as:
$J_{C_\pm}=\Pi_{C_\pm}{\Sigma_C}/{\Sigma}$~\cite{hill1966},
where $\Pi_{C_\pm}$ is the weight of the directed cycle $C_\pm$, $\Sigma_C$ is the sum of weights of spanning trees that are rooted on the cycle $C$, and $\Sigma$ is the sum of weights of spanning trees rooted on every individual state, a normalization factor identical for all cycles. Note that the weight of a subgraph is defined as the product of transition rates forming this subgraph. 
In practice, it will be challenging to count the vast number of spanning trees rooted on different vertices, especially when the quantum-transition network becomes complicated. Nevertheless, we surmount this difficulty by using the generalized matrix-tree theorem in algebraic graph theory, leading to a simple algebraic expression of cycle flux:
\begin{eqnarray}
J_{C_\pm}=\Pi_{C_\pm}\frac{\det(L[C;C])}{\sum_i\det(L[i;i])}.
\label{cycleflux}
\end{eqnarray}
$\det(L[C;C])$ is the principal minor of Laplacian $L$ of the weighted graph, obtained by deleting rows $i\in C$ and columns $i\in C$ from $L$. It equals the sum of weights of directed spanning trees rooted on the cycle $C$, i.e., $\det(L[C;C])=\Sigma_C$. Similarly, $\sum_i\det(L[i;i])=\Sigma$. 
This algebraic expression of cycle flux has an intuitive interpretation~\cite{jren2017fop} because the directed flow of weighted edges (transition rates) on spanning trees towards a cycle is associated with the occurring frequencies of the cycle trajectory. Alternative mathematical proof in terms of Markov chain can be found in Ref.~\cite{qianmin2004book}. 
In Fig.~\ref{fig1}(c), we show the four spanning trees rooted on cycle $C_{1\rightarrow2\rightarrow3\rightarrow1}$, the weight sum of which leads to the factor $\det(L[1,2,3;1,2,3])$. By multiplying the weight of the cycle itself $\Pi_{1\rightarrow2\rightarrow3\rightarrow1}=k_{1,2}k_{2,3}k_{3,1}$ upon normalizing a common factor $\sum_{i=1}^5\det(L[i;i])$, the cycle flux $J_{C_{1\rightarrow2\rightarrow3\rightarrow1}}$ can be readily calculated. Following the same procedure, we can calculate the cycle fluxes of all the cycle trajectories as listed in Fig.~\ref{fig1}(b), and rank out the dominant cycle fluxes. 

In what follows, we apply above cycle flux analysis to two typical transport models of quantum devices: one is a thermal-drag spin-Seebeck pump model, and the other one is a quantum thermal transistor model. Results demonstrate the advantages of the graph theory that the dominant cycles indeed elucidate their principal working mechanisms.

\begin{figure}[htb]
\includegraphics[scale=0.26]{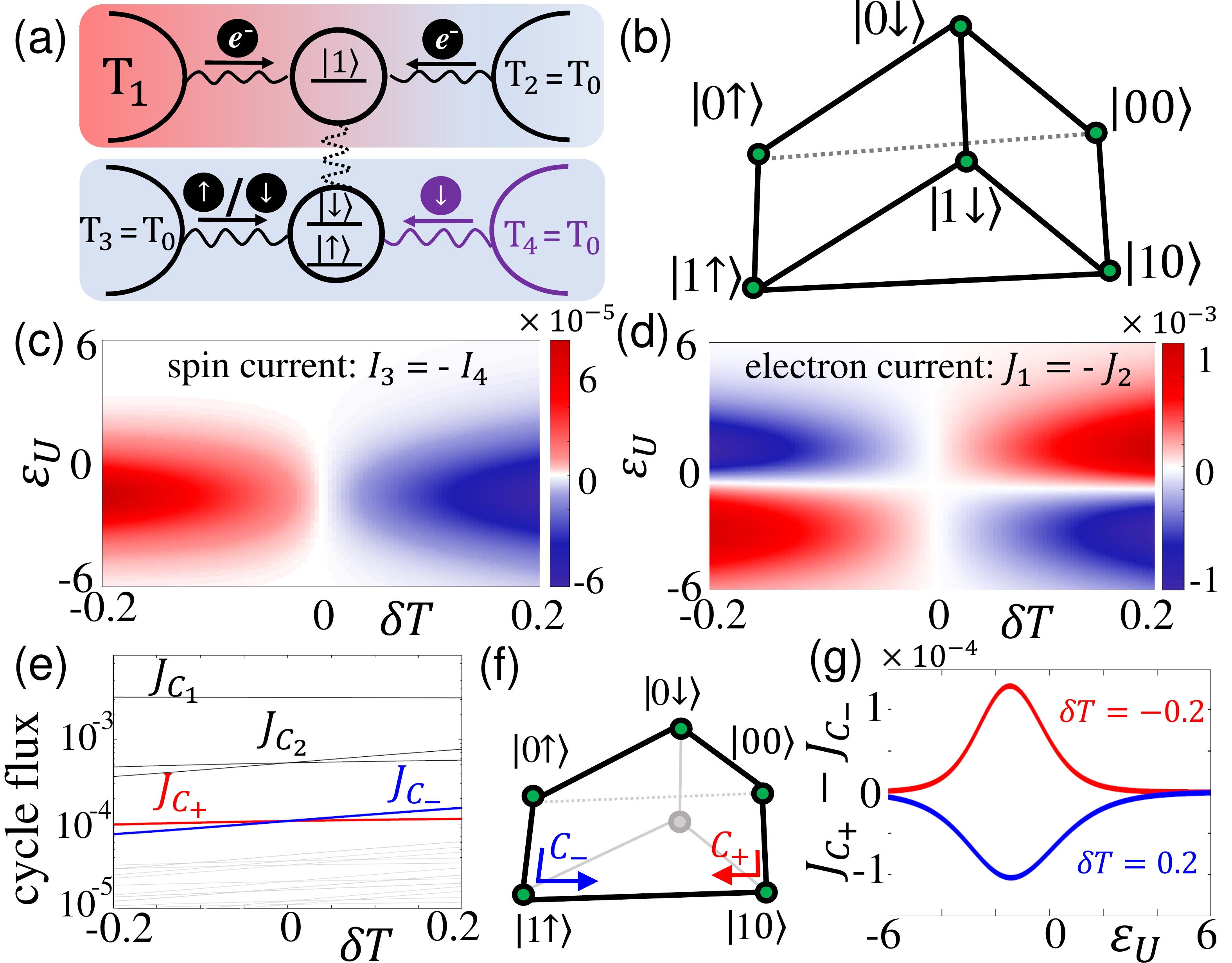}
\caption{\textbf{Cycle flux analysis of thermal-drag Spin-Seebeck pump.}
(a) Scheme of thermal-drag spin-Seebeck pump in hybrid dot structures. The upper dot is coupled with two spinless electron reservoir. The lower dot is coupled with a spinful electron reservoir (left) and a magnon bath (right). The two dots are Coulomb coupled. (b) Quantum transition network in state space.
(c) (d) Spin current $I_3$ (-$I_4$) and electron current $J_1$ (-$J_2$) as functions of $\delta T=T_1-T_0$ and $\varepsilon_U$, respectively, where $I_{i=3,4}$ ($J_{i=1,2}$) is the net spin (electron) current flowing from reservoir $i$ into the center system.
(e) Top ranked cycle fluxes at $\varepsilon_U=1$. The dominant cycle fluxes are represented by the bold red(light) and blue(dark) lines.
(f) The dominant paired cycle trajectories for thermal-drag spin current.
(g) The net cycle flux $J_{C_+}-J_{C_-}$ as a function of $\varepsilon_U$ for $\delta T=-0.2$ and $0.2$, respectively.
Other parameters are: $\varepsilon_{L\downarrow}=-\varepsilon_{L\uparrow}=1, U=3$, $T_2=T_3=T_4=T_0=1$ and calculation details are in supplementary section \uppercase\expandafter{\romannumeral2}~\cite{Suppl}  }~\label{fig2}
\end{figure}

\emph{i) Thermal-drag spin-Seebeck pump.} 
The first hybrid quantum model is illustrated in Fig.~\ref{fig2}(a). The upper dot, with two states $\ket{\phi_U}=\{\ket{0},\ket{1}\}$, is connected with two spinless electron reservoirs: $V_{U}=t_{U}\sum_{k,\nu=\{1,2\}} c_{U\nu,k}^{\dagger} d_{U} +{\rm H.c.}$.
While the lower dot, with three states $\ket{\phi_L}=\{\ket{0},\ket{\uparrow},\ket{\downarrow}\}$, is connected with both a left-side spinful electron reservoir and a right-side magnon bath~\cite{SSEprb2013}: $V_{L}=t_{L} \sum_{k,\sigma=\{\uparrow\downarrow\}} c_{L\sigma, k}^{\dagger} d_{L \sigma}+\gamma\sum_q b_q^\dagger d_{L\uparrow}^\dagger d_{L\downarrow}+{\rm H.c.}$.
$c_k^\dagger$ ($c_k$) and $b_q^\dagger$ ($b_q$) are creation (annihilation) operators in electron reservoir with momentum $k$ and in magnon bath with momentum $q$, respectively.
The upper and lower dots are Coulomb coupled with each other, with the Hamiltonian
\begin{eqnarray}
H_s=\sum_{v=U,L}\varepsilon_{v}\hat{n}_v+U\hat{n}_L\hat{n}_U,
\end{eqnarray}
where $\hat{n}_U=d_{U}^\dagger d_{U}$ and  $\hat{n}_L=\sum_{\sigma}d_{L \sigma}^\dagger d_{L \sigma}$ are the number operators of two dots. Although with the Coulomb energy $U$ coupling the upper and lower subsystems, the electron transport is only across the upper one while the spin transport occurs merely through the lower one. Under the eigenstate basis $\ket{\phi_U\phi_L}$, the corresponding quantum-transition network is depicted in Fig.~\ref{fig2}(b). 

An intriguing phenomenon emerges that a nonzero spin current across the lower isothermal subsystem is pumped by the nonzero thermal bias $\delta T=T_1-T_0$ across the upper subsystem, as shown in  Fig.~\ref{fig2}(c). 
By tuning $\varepsilon_U$, the thermal generated electron current in the upper subsystem will vanish and reverse due to the effective particle-hole symmetry, as shown in Fig.~\ref{fig2}(d). Yet, the thermal-excited spin Seebeck current in the lower subsystem still reaches maximum near $\varepsilon_U$ of vanishing electron current. This result provides a strong evidence suggesting mechanism beyond normal Coulomb drag effect caused by drifted electron current.

To further excavate the working cycle underlying the thermal spin pump effect, we decompose the network of spin-Seebeck process into a set of cycles and analyze the dominant cycles utilizing the cycle flux ranking. There are in total 28 directed cycles in Fig.~\ref{fig2}(b), and the corresponding top-ranked cycle fluxes are depicted in Fig.~\ref{fig2}(e). The top(second) bidirectional cycles $C_1$($C_2$) are futile (reason see supplementary section \uppercase\expandafter{\romannumeral2}~\cite{Suppl}). The third-ranked  paired cycles, i.e. $C_+$ and $C_-$,  dominate the spin current generation, of which the cycle trajectories are drawn in Fig.~\ref{fig2}(f). We describe the five steps of forward cycle $C_+$ as follows: Starting from $\ket{00}$, the system transits sequentially into $\ket{10}$ (one electron jumps into the upper dot), $\ket{1\uparrow}$ (one spin-up electron tunnels from reservoir $3$ into the lower dot, assisted by Coulomb repulsion), $\ket{0\uparrow}$ (the upper electron jumps out), $\ket{0\downarrow}$ (the lower spin-up electron flips to spin-down state by absorbing one magnon), and finally back to $\ket{00}$ (the lower spin-down electron jumps back to reservoir $3$).
Among them, $\ket{10}\rightarrow\ket{1\uparrow}$ and  $\ket{0\downarrow}\rightarrow\ket{00}$ together contribute a spin $1$ transfer from the left reservoir to the center system, while
$\ket{0\uparrow}\rightarrow\ket{0\downarrow}$ contributes the subsequent one spin transfer from the center part to the right bath. As a whole, the cycle $C_+$ completes an integer spin-1 transfer from the left electron reservoir to the right magnon bath. Similarly, the backward cycle $C_-$ gives the reversed process. 

In linear response regime, we can readily obtain the forward-backward ratio of the dominant cycle fluxes as:
\begin{eqnarray}~\label{eq5}
\frac{J_{C_{+}}}{J_{C_{-}}}=\frac{\Pi_{C_+}}{\Pi_{C_-}}\approx1-\frac{U\delta T}{2T_0^2},
\end{eqnarray}
where $\delta T=T_1-T_0$. This clearly demonstrates that the thermal bias ($\delta T$) in the upper subsystem and the Coulomb repulsion ($U$) between upper and lower subsystems synergistically determine the emergence of spin-Seebeck pump effect in the lower subsystem. Changing the sign of $\delta T$ would reverse the net dominant cycle flux $J_{C_+}-J_{C_-}$, as shown in Fig.~\ref{fig2}(g), which is consistent with the spin Seebeck current $I_3\propto J_{C_+}-J_{C_-}$, shown in Fig.~\ref{fig2}(c). For a given $\delta T$, changing $\varepsilon_U$ does not reverse the dominant cycle flux, although the electron current through the upper subsystem is reversed. Therefore, nonzero spin current generation (cycle flux $J_{C_+}-J_{C_-}$) at the zero electron current regime in the upper subsystem [shown in Fig.~\ref{fig2}(d)] clearly proves the working mechanism 
as a thermal-drag spin Seebeck pump effect, rather than the conventional electron-current drag effect.

\begin{figure}[htb]
\includegraphics[scale=0.14]{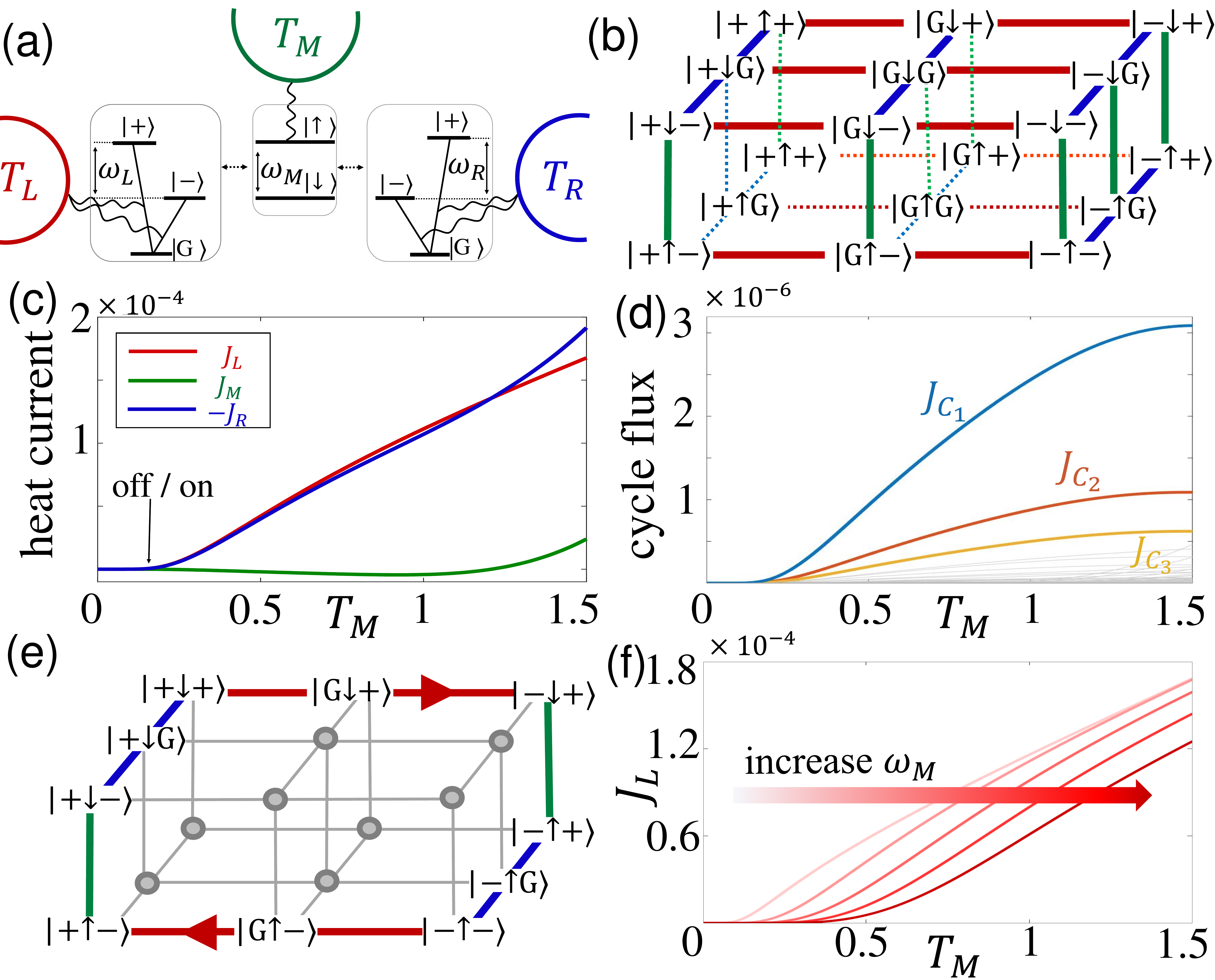}
\caption{\textbf{Cycle flux analysis of quantum thermal switch.}
(a) Schematic of the quantum thermal transistor, which is composed of qutrit-qubit-qutrit system, individually coupled to one thermal bath denoted by a half circle with $T_v$ $(v=L,M,R)$.
(b) Quantum transition graph in state space $\ket{LMR}$ with $\ket{L},\ket{R}=\ket{+},\ket{-},\ket{G}$ and $\ket{M}=\ket{\uparrow},\ket{\downarrow}$. Red, green and blue edges denote the state-transition in the left, middle and right parts, respectively.
(c) Heat current from left to right as a function of $T_M$.
(d) Top-3 ranked cycle fluxes versus $T_M$.
(e) The dominant cycle $C_1$ embedding in the state network.
(f) Increasing the switch threshold by increasing $\omega_M$ from 0.5 to 2.5 with interval 0.5.
Parameters are
$\omega_{L,M,R}=1$, $\omega_{LM}=\omega_{MR}=10$, $\omega_{LR}=0$, $\omega_{0,L(R)}=3$, $T_L=2.5$, $T_R=0.2$ and calculation detail see supplementary section \uppercase\expandafter{\romannumeral3}~\cite{Suppl}. 
}~\label{fig3}
\end{figure}

\emph{ii) Quantum thermal transistor.} We then apply the cycle flux analysis to investigate typical functionalities of the second quantum device model, such as quantum thermal switch, heat amplifier and NDTC.
Figure~\ref{fig3}(a) shows a hybrid qutrit-qubit-qutrit system connected to three individual boson baths with the Hamiltonian $H_{b,v=\{L,M,R\}}=\sum_k\omega_{v,k}a^\dagger_{v,k}a_{v,k}$. The left (right) \emph{V}-type qutrit is coupled with the photon bath via  $V_{u=\{L,R\}}=\sum_k(g_{u,k}^+a_{u,k}^\dagger\ket{G_u}\bra{ +_u}+g_{u,k}^-a_{u,k}^\dagger\ket{G_u}\bra {-_u})+{\rm H.c.}$, and the middle one is coupled with a phonon bath via $V_M=\sum_kg_{M,k}(a_{M,k}^\dagger\sigma_M^-+a_{M,k}\sigma_M^+)$, where $g_{u,k}$ is the coupling strength. The hybrid central Hamiltonian takes the variant of Ref.~\cite{kjoulain2016prl} as
\begin{eqnarray} \label{eq7}
&&H_s=\sum_{u=L,M,R}\frac{\omega_u}{2}\sigma^z_u+\sum_{v=L,R}(\frac{\omega_{v\scaleto{M}{3.5pt}}}{2}\sigma^z_v\sigma^z_M+\omega_{\scaleto{0}{3pt},v}\hat{d}_v),
\end{eqnarray}
where $\sigma^z_{M}$ is the middle qubit Pauli martrix, and the left and right qutrits are $\sigma^z_{v}=\ket{+_v}\bra{+_v}-\ket{-_v}\bra{-_v}$ and $\hat{d}_v=\ket{+_v}\bra{+_v}+\ket{-_v}\bra{-_v}$, where $(v=L, R)$. The quantum-transition network of eigenstates of the hybrid system is illustrated in Fig.~\ref{fig3}(b), which has a highly symmetric topology.
Figure~\ref{fig3}(c) illustrates the quantum thermal switch effect by tuning the middle gating temperature $T_M$. 
When $T_M$ is sufficiently low, three heat currents are suppressed and the system is at the \emph{off} state. 
By continuously increasing $T_M$ beyond a threshold value, the source and drain heat current $J_L$ and $-J_R$ are dramatically enhanced, with still negligible gating heat current $J_M$~\cite{bkarimi2017qst}, 
As such, the system switches to the \emph{on} state.  
We enumerate all the cycle trajectories, and calculate the  cycle fluxes based on Eq.~(\ref{cycleflux}), 
to dissect which process dominates the switch behavior in this quantum thermal transistor. 

We plot top-3 ranked cycle fluxes in Fig.~\ref{fig3}(d). 
Obviously, the flux of cycle $C_1$, as marked in Fig.~\ref{fig3}(e), dominates the heat current, 
which presents similar profile and closed switch point ($T_M \approx 0.2$) to those of $J_{L(R)}$.
As the principal working cycle, $C_1$ is indispensable to $J_L$ and $J_R$, but irrelevant with $J_M$. 
The reason is that within $C_1$, the net energies transferring along the middle-bath-involved trajectory $\ket{-\downarrow +}\rightarrow\ket{-\uparrow +}$ and $\ket{+\uparrow -}\rightarrow\ket{+\downarrow -}$ are cancelled to zero, {\it i.e.}, $(\omega_{\ket{-\downarrow +}}-\omega_{\ket{-\uparrow +}})+(\omega_{\ket{+\uparrow -}}-\omega_{\ket{+\downarrow -}})=0$. 
Therefore,  during this cycle there is no heat exchange between the hybrid system and the middle thermal bath, 
and the heat current flows from the left bath equals  the heat current flows into the right bath.

Further dissecting $J_{C_1}$, we find the transition rate $k_{\ket{-\downarrow +}\rightarrow \ket{-\uparrow +}}\propto 1/(e^{\omega_M/T_M}-1)$ in $\Pi_{C_1}$ [see Eq.~(\ref{cycleflux})] governs the switch threshold by the ratio $\omega_M/T_M$. 
Once the gating temperature $T_M>\omega_M/5$, the middle qubit will be significantly thermal excited to open the transport channel from left to right. 
As such, the device switches from $off$ state to $on$ state. From the cycle flux analysis, we know that the switch effect relies on a principal factor $\omega_M/T_M$, which guides us to inverse design the quantum thermal switch with tunable threshold. 
Indeed, by increasing $\omega_M$ in Fig.~\ref{fig3}(f), 
the gating temperature for switching increases to larger threshold $T_M$. 
Cycle trajectories $C_2$ and $C_3$ hold similar behaviors as $C_1$, of which relevant details and discussions can be found in supplementary section \uppercase\expandafter{\romannumeral3} ~\cite{Suppl}.

\begin{figure}[htb]
\includegraphics[scale=0.25]{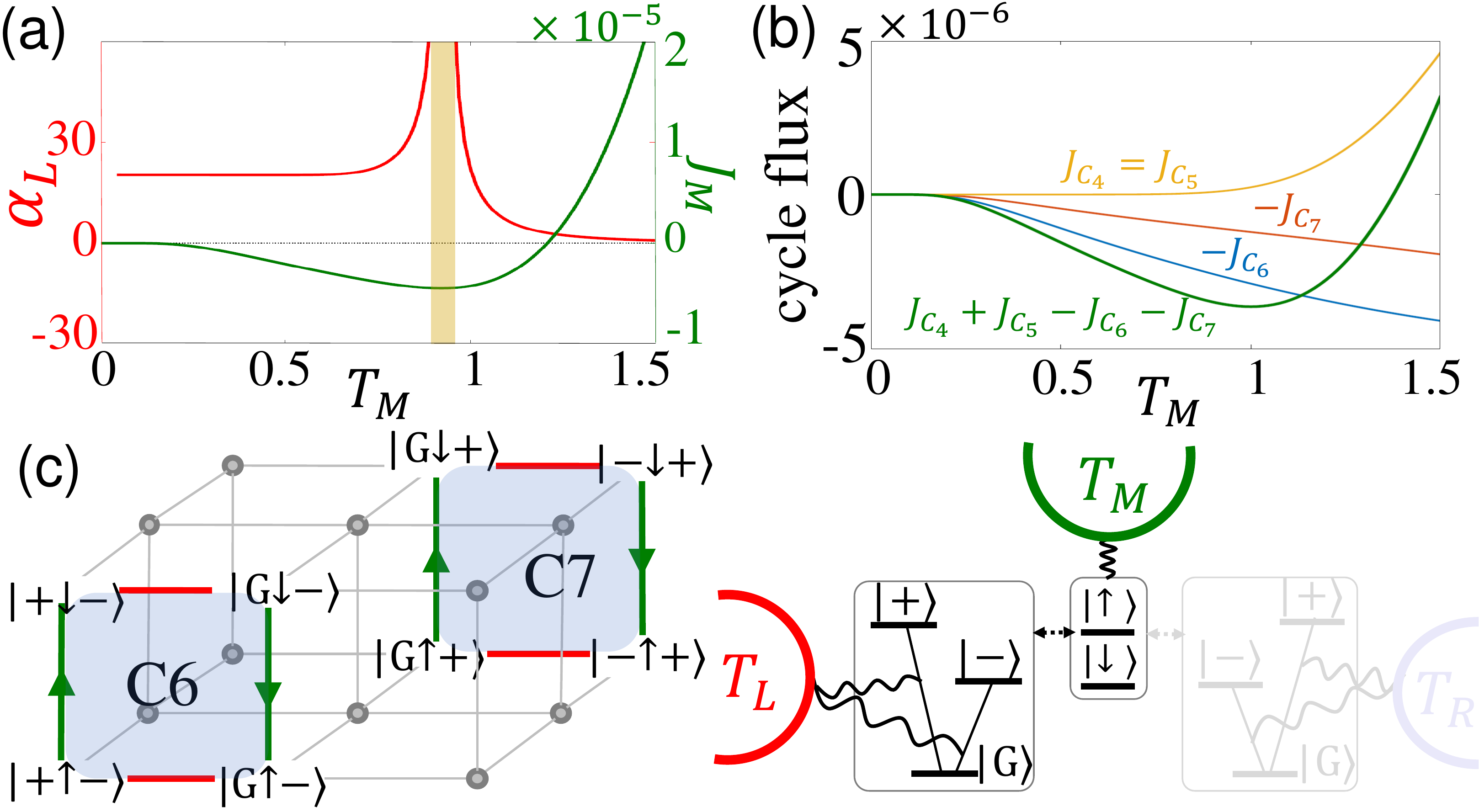}
\caption{\textbf{Cycle flux analysis of heat amplification and NDTC.}
(a) Amplification factor $\alpha_L$ and $J_M$ as a function of gating temperature $T_M$. Parameters are the same as in Fig.~\ref{fig3}.
(b) The fluxes of four cycles $C_4, C_5, C_6, C_7$ dominate the contribution to $J_M$. 
The sign ``-" before $C_6$ and $C_7$ means the negative contribution, leading to the NDTC effect. 
(c) Cycle trajectories $C_6$ and $C_7$ responsible for the NDTC, corresponding to a reduced working mechanism with the right qutrit being frozen.
}
~\label{fig4}
\end{figure}

Similar to the negative differential electric conductance observed in the p-n junction by L. Esaki~\cite{lesaki1958pr}, the NDTC in heat transport~\cite{rj2013prbndtc} is also a crucial ingredient of heat amplification~\cite{bli2006apl} and thermal logic gates~\cite{lwang2007prl}.
Figure~\ref{fig4}(a) shows that in a broad range of $T_M$, the system exhibits a stable amplification factor: $\alpha_L=|\partial J_L/\partial J_M|\approx 20$, where the counterintuitive NDTC appears: the heat current flowing into the middle thermal reservoir ($-J_M$) anomalously increases as this reservoir temperature increases: $\partial{(-J_{M})}/\partial{T_M}>0$. A giant amplification factor diverges around the yellow regime, where $J_M$ becomes flat ($\partial J_M/\partial T_M \rightarrow0$) at the crossover regime between negative and positive differential thermal conductance.

Looking into the cycle fluxes listed in Fig.~\ref{fig4}(b), we identify that the cycle flux of $C_6$($C_7$) dominates the NDTC of $J_M$, flowing from the middle reservoir into the system, of which the cycle trajectories are illustrated in Fig.~\ref{fig4}(c). Interestingly, during the whole cycle trajectory $C_6$($C_7$) the right qutrit keeps frozen at state $\ket{-}$($\ket{+}$), which describes a simplified working picture that the reduced system absorbs energy from the left reservoir and emits energy into the middle bath by qubit flipping, contributing negative to $J_M$.  By increasing $T_M$, although the thermal bias $T_L-T_M$ becomes smaller, the middle qubit flipping is easier thermally excited by the middle reservoir so that the transport channel is dramatically enhanced, leading to the NDTC effect $\partial{J_{M}}/\partial{T_M}<0$. 

Higher $T_M$ also opens new transport channels, i.e., cycles $C_4$ and $C_5$ (trajectories are in supplementary \uppercase\expandafter{\romannumeral3}~\cite{Suppl}), which emit energy into the middle reservoir from the system and contribute positive to $J_M$. Competing with the NDTC cycles $C_6$ and $C_7$, the faster increasing $J_{C_4}$ and $J_{C_5}$ [see Fig.~\ref{fig4}(b)] induces 
the non-monotonicity of $J_M$, leading to the giant heat amplification at the inflexion point.
The cycle flux ranking of network analysis indeed help us to grasp the principal working mechanism of quantum thermal devices out of their complex transport behaviors in different parameter range.

\emph{Conclusion}.
By mapping the dissipative quantum devices into networks of quantum state transitions, we have decomposed the transition network into cycle trajectories, and drawn a systematic connection between the cycle graph theory and nonequilibrium quantum transport. 
The cycle flux ranking of network analysis is introduced to analyze typical quantum thermal devices.
With the help of algebraic graph theory, we have demonstrated that the dominant cycle fluxes genuinely grasp the principal mechanism out of complex transport features, such as thermal-drag spin-Seebeck pump, quantum thermal switch, heat amplification, and negative differential thermal conductance. Cycle flux analysis provides researchers an alternative viewpoint to investigate quantum transport and would promote the inverse design and optimization of multi-functional quantum devices.

\emph{Acknowledgements}.
 The work is supported by the National Natural Science Foundation of China (Nos. 11935010, 11775159, and 11704093), and the Opening Project of Shanghai Key Laboratory of Special Artificial Microstructure Materials and Technology.

\bibliography{ref}

\end{document}